\documentclass[%
reprint,
superscriptaddress,
 amsmath,amssymb,
 aps,
]{revtex4-2}

\usepackage{graphicx}% Include figure files
\usepackage{dcolumn}% Align table columns on decimal point
\usepackage{bm}% bold math
\usepackage{multirow}
\usepackage[ruled,vlined]{algorithm2e}
\newcommand{\ignore}[1]{}
\usepackage{lineno}
\usepackage{color}

\usepackage{amsmath} % using math package
\usepackage{amssymb}
\usepackage{multirow} %表格中单元格跨行
\usepackage{xcolor}
\newcommand{\ket}[1]{\big|  #1 \big \rangle }
\newcommand{\bra}[1]{ \big\langle #1 \big | }

\makeatletter %使\section中的内容左对齐
\renewcommand{\section}{\@startsection{section}{1}{0mm}
{-\baselineskip}{0.5\baselineskip}{\bf\leftline}}
\renewcommand{\subsection}{\@startsection{subsection}{1}{0mm}
{-\baselineskip}{0.5\baselineskip}{\bf\leftline}}
\makeatother

\begin{document}

\title{Finding efficient observable operators in entanglement detection via convolutional neural network}% Force line breaks with \\

%\author{Authors}
%\affiliation{Affiliations}

%\iffalse

\author{Zi-Qi Lian}
%\thanks{Liu-Jun Wang and Kai-Yi Zhang contribute equally to this work.}
\affiliation{School of Physics and Astronomy and Yunnan Key Laboratory for Quantum Information, Yunnan University, Kunming 650500, China}
\author{You-Yang Zhou}
\affiliation{School of Physics and Astronomy and Yunnan Key Laboratory for Quantum Information, Yunnan University, Kunming 650500, China}
\author{Liu-Jun Wang}
\email{ljwangq@ynu.edu.cn}
\affiliation{School of Physics and Astronomy and Yunnan Key Laboratory for Quantum Information, Yunnan University, Kunming 650500, China}
\author{Qing Chen}
\email{chenqing@ynu.edu.cn}
\affiliation{School of Physics and Astronomy and Yunnan Key Laboratory for Quantum Information, Yunnan University, Kunming 650500, China}

%\fi


%\section*{\label{Abstract}Abstract}
\begin{abstract}
In quantum information, it is of high importance to efficiently detect entanglement. Generally, it needs quantum tomography to obtain state density matrix. However, it would consumes a lot of measurement resources, and the key is how to reduce the consumption. In this paper, we discovered the relationship between convolutional layer of artificial neural network and the average value of an observable operator in quantum mechanics. Then we devise a branching convolutional neural network which can be applied  to detect entanglement in 2-qubit quantum system. Here, we detect the entanglement of
%\textcolor{red}{}
Werner state, generalized Werner state and general 2-qubit states,
and observable operators which are appropriate for detection can be automatically found.
%\textcolor{red}{}
Beside, compared with privious works, our method can achieve higher accuracy with fewer measurements for quantum states with specific form.
The results show that the convolutional neural network is very useful for efficiently detecting quantum entanglement. 
\end{abstract}

\maketitle

%\tableofcontents

\section*{\label{Introduction}Introduction}
%\subsection*{\label{Introduction}Introduction}

Nowadays, Machine Learning has become a powerful tool in tackling some complicated quantum physics problems, beacuse of its ability to find potential patterns in vast data. The breakthroughs have been made in multi-particle quantum system state ansatz \cite{S1-RBMstate,S2-NNtomography,S3-Purification-via-Neural-Density,S4-Eigenstate-extraction,S5-Reconstructing-quantum-states,S8-En-in-Deep-Learning,S9-CNNstate1,S10-CNNstate2,S44-NNdensity,S11-Phases-of-spinless-lattice,S12-HybridCNN,S23-Entanglement-in-NNS,G8-Efficient-representation-state}, discovering quantum phase transition \cite{LZXB-1,LZXB-2,LZXB-3,LZXB-4,LZXB-5,LZXB-6}, classifying quantum correlations \cite{G24-Ma,G25-sample-convex-hull,G26,G27-from-disorder-system,G28,G48-find-nonlocality,G49-branching-FC,G52-quantum-discord,G53-Ren-multiple-classify,G54-Ren-Steerability-detection}, detecting entanglement structure \cite{G29-Entanglement-Structure} and estimating the violation of multi-particle Bell inequalities \cite{G47-bellinequality}, etc. On the other hand, with the development of the quantum computer, scientists pay more attention to quantum Machine Learning algorithms, which will be implemented on quantum computer, such as quantum approximate optimization algorithm \cite{S56-QAOA}, variational quantum eigensolver \cite{S57-eigenvalue-solver}, quantum Boltzmann machine \cite{S59-QRBM}, and quantum neural network \cite{S47-QCNN,S58}. They will promote the development of Machine Learning.

Among complicated quantum physics problems, the detection of quantum entanglement is an essential one. Quantum entanglement is the essential resource in application of quantum teleportation \cite{K55-quantum-teleportation}, quantum key distribution \cite{K57-keydistribution} and quantum computation \cite{K56-Quantum-Compute}. %There is an explicit mathematical definition of quantum entanglement\cite{Entanglement}.
Although there are many entanglement detection criteria have been proposed such as positive partial transpose (PPT) criterion \cite{PPTcriterion1,PPTcriterion2}, computable cross norm criterion or realignment \cite{CCNR1,CCNR2} and entanglement witnesses \cite{Witness1,Witness2,Witness3,Witness4,Witness5}  etc, %Most of them always require complete state information to reconstruct density matrix, which means lots of measurements in experiment. 
complete classifying entanglement is still an NP-hard problem \cite{NPhard}. %(xiugai?!)

%(MaYueChi: ..., a major challenge is to look for an efficient means for classifying quantum states)

%(G27: ...Nevertheless, the measurement of entanglement for many particle systems is very challenging.)

%(G25:...Despite the many methods—such as the positive partial transpose criterion and the k-symmetric extendibility criterion—to tackle this problem, none of them enables a general, practical solution due to the problem’s NP-hard complexity. Explicitly, separable states form a high-dimensional convex set of vastly complicated structures. ... In fact, such an entanglement detection problem is proved to be NP hard [6], implying that it is almost impossible to devise an efficient algorithm in complete generality. ... )

%(G31:...The task of classifying the entanglement properties of a multipartite quantum state poses a remarkable challenge due to the exponentially increasing number of ways in which quantum systems can share quantum correlations. ... A key instance of problems where ANN-based approaches hold the promises for a game-changing contribution is the discrimination of entangled and separable states, which is a known NP-hard classification problem in quantum information processing)

Recent years, scientists have also done many researches on using machine learning to study quantum entanglement. For instance, based on deterministic measurement operators, artificial neural network can be used to classify the entanglement in 2-qubits or 3-qubits systems \cite{G24-Ma,G53-Ren-multiple-classify}. Combined with supervised learning, convex hull approximation can sample a mass of separable pure state to approximate the shape of separable states set \cite{G25-sample-convex-hull}. Convolutional neural network (CNN) can estimate the entanglement entropy of disordered interacting quantum many-particle system \cite{G27-from-disorder-system}. And fully connected neural network (FC) can be used to predict the multipartite entanglement structure of states composed by random subsystems \cite{G29-Entanglement-Structure} or find semi-optimal measurements for entanglement detection \cite{2020Finding}.

In this work, we successfully applied CNN to detect entanglement. CNN is one of the most representative neural networks which is considered more efficient than FC \cite{K22-CNNmore-efficient}. At present, CNN has been used to express quantum state of multi-particle system \cite{S9-CNNstate1,S10-CNNstate2,S11-Phases-of-spinless-lattice,S44-NNdensity}, estimate entanglement entropy of quantum many-particle system \cite{G27-from-disorder-system} and the parameters of multi-particle Hamiltonian \cite{S40-CNN-Hamiltonian}. 
%\textcolor{red}{}
Here, we first show why the observable operator of quantum system with discrete energy levels can be regarded as a special convolution kernel and how to use a convolution kernel to represent an observable operator and get the average of the operators.
Afterward, we prove that the Hermiticity of convolutional layers can maintain in the training course if the input and the kernels are initialized as Hermitian matrix.

Then, we devised branching convolutional neural network (BCNN) (depicted in Fig.\ref{convolutional_pathes_and_BCNN}(c)) which has a number of independent convolutional pathes. Every convolutional path accurately calculates average value of observable operators in inputted quantum state. According to the features of quantum state, the structures of convolutional pathes, it can automatically find appropriate observable operators whitch can extract information required by training goal. 
%\textcolor{red}{}
Because of that, it can decrease resource consumption in practice.
We detected the entanglement of 2-qubits state and research the influence of the number of observable operators on the accuracy of our model.

\section*{Results}
%\subsection*{Encoding observable operators into convolutional kernels}
\subsection*{Regard observable operator as a kernel}

%(Inspired by that CNN extracts features by kernels, %we adopt it (dipicted in Fig.\ref{branching convolutional neural network}),
%we propose BCNN (dipicted in Fig.\ref{convolutional_pathes_and_BCNN}(c)) to extract entanglement information from quantum state.  
%(jie shao CNN)
%(shi fou bu yong zhe yang de zong jie)In this section, we will show that the convolutional layer can accurately calculate the average value of observable operator. We take state density matrix $\rho$ as the input of the neural network, and encode the transpose of observable operator $M^T$ into the kernel. If the step size of the kernel is exactly equal to its own dimension, then the output of the convolutional layer without activation function and bias is precisely equal to the average value of the observable operator $M$ for the quantum state $\rho$. It also can be proved that the gradient of the kernel is also Hermitian. It means the optimization methods based on gradient descent are still available. )

\begin{figure}
	\centering
	\includegraphics[width=3.25in]{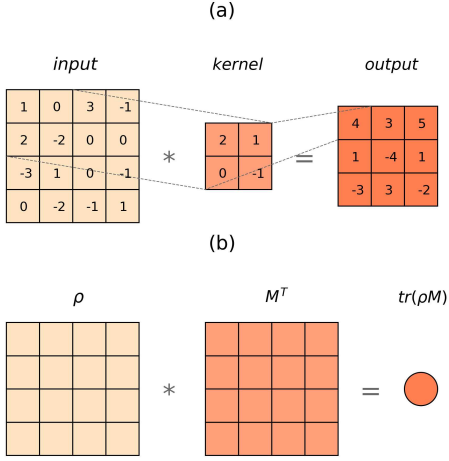}
	\caption{\label{simple example and convolutional path0}
		\textbf{(a),} A example of convolution without bias and activation function. The step size of kernel equal to 1. \textbf{(b),}The convolutional layer with the input $\rho$ and the kernel $M^T$, and its output is $tr(\rho M)$.}
\end{figure}

%(brief introduction of convolution:) 
CNN extract features from input data by the kernels, whitch are composed of trainable parameters. Every kernel scans the input data according to a certain step size. In each step, its parameters are multiplied with the corresponding input data and all are added as output. We can see a simple example of convolution without bias and activation function in Fig.\ref{simple example and convolutional path0}(a).

As we know, quantum states can completely describe a system. Observable operators can extract features such as momentum, spin, position, etc. from quantum states. From the point of feature extraction, we prove that the observable operator of discrete level system is a special convolution kernel. In Fig.\ref{simple example and convolutional path0}(b), we show how the convolutional layer can accurately calculate the average value of observable operator. We take state density matrix $\rho$ as the input of the neural network and the transpose of observable operator $M^T$ as the kernel, and the output of convolution without activation function and bias is

\begin{equation}\label{eq:1}
	\begin{split}
		\rho\ast M^T &=\sum_{ij}{\bra{i}\left( \sum_{kl}{\rho_{kl}M_{lk}\ket{k}\bra{l}}\right)\ket{j} }\\
		&=\sum_{ij}{\rho_{ij}M_{ji}}\\
		&=\langle M\rangle,
	\end{split}
\end{equation}
where $\ast$ means the convolution in the artificial neural network. If there are $2$ subsystem with the dimension $d^{(1)}$ and $d^{(2)}$, its state density matrix can be written as $\rho=\sum_{ij,kl}^{d^{(1)},d^{(2)}}{\rho_{ijkl}\ket{i}\bra{j}\otimes \ket{k}\bra{l} }$ and the observable operator $M$ can be written as $M=M^{(1)}\otimes M^{(2)}$. In the same way, we take $M^{(2)T}$ and $M^{(1)T}$ as the kernels of the first and the second convolutional layers, and their step sizes are exactly equal to their own dimensions. The output of the first convolutional layer is

\begin{figure*}
	\centering
	\includegraphics[width=6.5in]{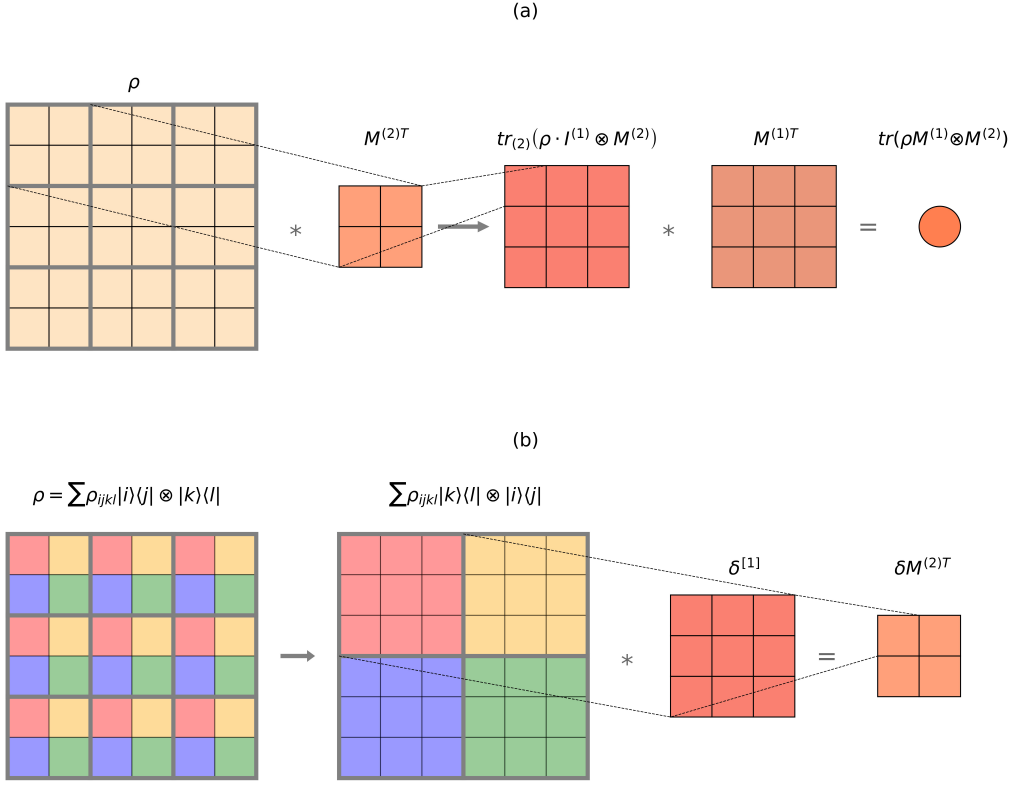}
	\caption{\label{convolutional path1}Take the $3\times 2$ system as an example. \textbf{(a),} There are two convolutional layers without activation function and bias. The first layer has the kerner $M^{(2)T}$ and the input $\rho$. The second has the kernel $M^{(1)T}$. The output of the second convolutional layer is exactly equal to $\langle M^{(1)}\otimes M^{(2)}\rangle$.
	\textbf{(b),} The gradient $\delta M^{(2)T}$ of the kernel $M^{(2)T}$ can be obtained by calculating convolution of $\sum{ \rho_{ijkl}|k\rangle\langle l|\otimes |i\rangle\langle j|}$ and $\delta^{[1]}$, where $\sum{ \rho_{ijkl}|k\rangle\langle l|\otimes |i\rangle\langle j|}$ can be obtained by swapping the corner labels of $\rho$ and $\delta^{[1]}$ is the error propagated into the first convolutional layer.}
\end{figure*}

\begin{equation}\label{eq:2}
	\begin{split}
		O^{[1]}&=\rho\ast M^{(2)T}\\
		&=\left(\sum_{ij,kl}^{d^{(1)},d^{(2)}}{ \rho_{ij,kl}\ket{i}\bra{j}\otimes \ket{k}\bra{l} }\right)\ast M^{(2)T} \\
		&=\sum_{ij,kl}^{d^{(1)},d^{(2)}}{ \rho_{ij,kl}\ket{i}\bra{j}\otimes\left( \ket{k}\bra{l}\ast M^{(2)T}\right) } \\
		&=\sum_{ij}^{d^{(1)}}\sum_{kl}^{d^{(2)}}{ \rho_{ij,kl}\bra{l}M^{(2)}\ket{k}\ket{i}\bra{j} } \\
		&=tr_{(2)}\left(\rho\cdot I^{(1)}\otimes M^{(2)} \right).
	\end{split}
\end{equation}
i.e., the first convolutional layer calculate the partial trace for the subsystem $(2)$ of $\rho\cdot I^{(1)}\otimes M^{(2)}$, thus $O^{[1]}$ is Hermitian and its dimension is $d^{(1)}$. Then, the output of the second convolutional layer can be got as

\begin{equation}\label{eq:3}
	\begin{split}
		O^{[2]}&=O^{[1]}\ast M^{(1)T}\\
		&=\left(\rho\ast M^{(2)T}\right)\ast M^{(1)T} \\
		&=tr_{(2)}\left(\rho\cdot I^{(1)}\otimes M^{(2)} \right) \ast M^{(1)T} \\
		&=tr\left[ \rho\cdot\left(M^{(1)}\otimes M^{(2)} \right) \right] \\
		&=\langle M^{(1)}\otimes M^{(2)} \rangle.
	\end{split}
\end{equation}

Similarly, suppose that there are $N$ subsystems with dimension $d^{(1)}, d^{(2)},\cdots, d^{(N)}$, and the observable operator $M$ can be written as $M=M^{(1)}\otimes M^{(2)}\otimes\cdots \otimes M^{(N)}$, it is possible to caculate its average via the convolutional path with $N$ convolutional layers. For $\forall n\leq N$, the kernel of the $n$-th convolutional layer is $M^{(N-n+1)T}$, and the step size equal to its dimensions. Therefore, for $\forall n< N$, the output $O^{[n]}$ is
\begin{equation}\label{eq:4}
	\begin{aligned}
		O^{[n]}&=O^{[n-1]}\ast M^{(N-n+1)T}\\
		&=tr_{(N-n+1)}(O^{[n-1]}\cdot I^{(1)}\otimes\cdots\otimes I^{(N-n)} \\
		& \otimes M^{(N-n+1)} )\\
		&=tr_{(N-n+1,\cdots,N)} (\rho I^{(1)}\otimes \cdots \otimes I^{(N-n)}\\
		& \otimes M^{(N-n+1)}\otimes\cdots \otimes M^{(N)}) .
	\end{aligned}
\end{equation}
Likewise, it can be prove that $O^{[n]}$ is also Hermitian, and its dimension is $d^{(O)}= d^{(1)}\cdot d^{(2)}\cdots d^{(N-n)}=d^{(O')}\cdot d^{(M')}$, where $d^{(O')}$ and $d^{(M')}$ are the dimensions of the output $O^{(n+1)}$ and kernel $M^{(N-n)T}$ of next layer. In the same way, the output of the last layer also can be obtained 

\begin{equation}\label{eq:5}
	\begin{split}
		O^{[N]}=&\left(\left(\left(\rho \ast {_{}^{}}{M}{_{}^{(N)T}} \right)\ast {_{}^{}}{M}{_{}^{(N-1)T}}\right)\ast\cdots\right)\ast{_{}^{}}{M}{_{}^{(1)T}}\\
		=&\langle {_{}^{}}{M}{_{}^{(1)}} \otimes {_{}^{}}{M}{_{}^{(2)}}\otimes \cdots \otimes {_{}^{}}{M}{_{}^{(N)}} \rangle .
	\end{split}
\end{equation}

Furthermore, considering that artificial neural networks are usually trained  based on gradient descent, we  prove that if the input and kernel of the convolutional layer are initialized as Hermitian matrixes, the gradient of the kernel will also be Hermitian. The calculation of kernel's gradient depends on the neural network error matrix from back propagation and the input of convolutional layer. We let $\delta^{[n]}$ be the error matrix whtich is propagated into the $n$-th convolutional layer. Its dimension always equal to dimension of $O^{[n]}$. Since the step size of the kernels here are equal to their dimensions, for $\forall n< N$, $\delta^{[n]}$ can be expressed as 
\begin{equation}\label{eq:6}
	\begin{split}
		\delta^{[n]}=\delta^{[n+1]} \otimes M^{(N-n)T}.
	\end{split}
\end{equation}

Because of the Hermitianity of $O^{[N]}$, the error $\delta^{[N]}$, which is propagated from the fully connected layer, must be a real number too. It means for $\forall n \leq N$, $\delta^{[n]}$ is Hermitian. Considering that, for $\forall n < N$, $d^{(O)}=d^{(O')}\cdot d^{(M')}$, so $O^{[n]}$ can be written as $\sum_{ij,kl}^{d^{(O')},d^{(M')}}{O_{ijkl}^{[n]}\ket{i}\bra{j}\otimes\ket{k}\bra{l}}$. Then according to the neural network back propagation theory, for $\forall n \leq N$, the kernel gradient is 
\begin{equation}\label{eq:7}
	\begin{split}
		\delta M^{(N-n+1)T}&=\sum_{ij,kl}^{d^{(O')},d^{(M')}}{O_{ij,kl}^{[n-1]} \ket{k}\bra{l}\otimes\ket{i}\bra{j}}\ast\delta^{[n]} \\
		&=\sum_{kl}^{d^{(M')}}\sum_{ij}^{d^{(O')}}{O_{ij,kl}^{[n-1]} \bra{j}\delta^{[n]T} \ket{i} \ket{k}\bra{l}}.
	\end{split}
\end{equation}
Since $O^{[n-1]}$ and $\delta^{[n]}$ are Hermitian, so $\sum_{ij}{O_{ij,kl}^{[n-1]} \bra{j}\delta^{[n]T} \ket{i}}=( \sum_{ij}{O_{ij,lk}^{[n-1]} \bra{j}\delta^{[n]T} \ket{i}} )^\ast$, as well as, $\delta M^{(N-n+1)T}$ is also Hermitian. Therefore, in the process of updating based on gradient descent, the Hermitianity of the kernel will not change. 

So far, we prove that the observable operator of the discrete level system can indeed be regarded as a special convolution kernel,  the convolutional layer can be used to calculate the average of observable operators, and these convolutional layers can naturally keep Hermitianity when trained by gradient-based optimization methods.

\subsection*{Entanglement detection for 2-qubits state}

\begin{figure*}
	\centering
	\includegraphics[width=6.5in]{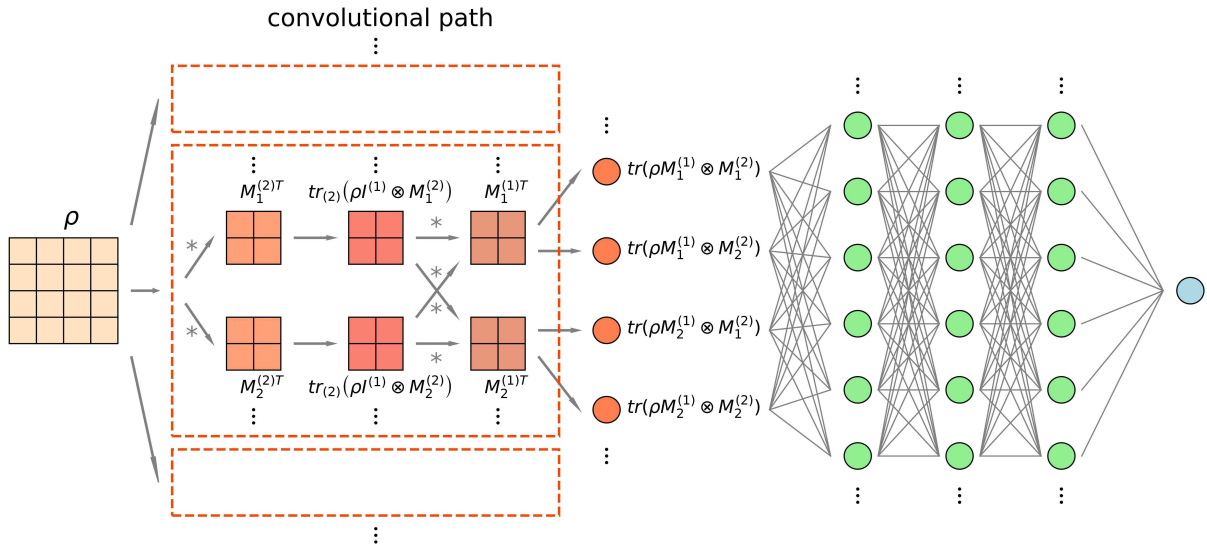}
	\caption{\label{convolutional_pathes_and_BCNN} Branching convolutional neural network(BCNN). The input of the network is the density matrix $\rho$, and it goes through several independent convolutional paths  (showed in red dotted box). Every convolutional path has two convolutional layers and each convolutional layer has several kernels. Each convolutional path outputs the average of all combinations of the kernels of its two convolutional layers. Then, we take the outputs of these convolutional paths as the input of fully connected layer to classify the entanglement of states. For Werner,  G\uppercase\expandafter{\romannumeral1}-Werner and G\uppercase\expandafter{\romannumeral2}-Werner states, we use three fully connected layers. For general 2-qubits state, we add two fully connected layers in the former structure.}
\end{figure*}

Based on the content we introduced above, we devise the BCNN (depicted in Fig.\ref{convolutional_pathes_and_BCNN}) to classify the entanglement of 2-qubits state. BCNN consists of several convolution paths and the following fully connected layers. It can automatically find proper observable operators which can extract information needed for the training goal. Here, we use $(m; n_1, n_2)$ to describe the structure of convolutional paths, where $m$ means how many convolutional pathes the network has, $n_1$ and $n_2$ means there are two layers of convolutional layer in a convolutional path and they have $n_1$ and $n_2$ kernels respectively . After training, the trained observable operators can be obtained from these kernels. More details of BCNN is introduced in the section Methods\ref{Methods}. Our dataset consists of state density matrixes and corresponding entanglement labels. The labels are determinded by PPT criterion, which is necessary and sufficient for entanglement classification of $2\times 2$ and $2\times 3$ system \cite{necessary-sufficient}. Next, we will briefly introduce the quantum states we tested. The Werner state is 
\begin{equation}\label{eq:8}
	\rho=p \ket{\psi} \bra{\psi} + \frac{ \left( 1-p\right) I}{4},
\end{equation}
where, $\ket{\psi}=\frac{1}{\sqrt{2}} \left( \ket{00}+\ket{11} \right)$, $p\in (0,1)$. It has only one free parameter $p$, when $p> \frac{1}{3}$ it is entangled \cite{Entanglement}. 

The first generalized Werner state which we called G\uppercase\expandafter{\romannumeral1}-Werner state is
\begin{equation}\label{eq:9}
	\rho(\theta)=p\ket{\psi _\theta} \bra{\psi _\theta}+\frac{(1-p)I_A }{2} \otimes \rho_{B}^{\theta},
\end{equation}
where, $\ket{\psi _\theta}=\cos\theta \ket{00}+\sin\theta \ket{11}$, $p\in (0,1)$, $\theta \in (0,2\pi)$, and ${_{}^{}}{\rho}{_{B}^{\theta}}=tr_A(\ket{\psi _\theta} \bra{\psi _\theta})$ is the reduced density matrix of the B system. The G\uppercase\expandafter{\romannumeral1}-Werner state has two free parameters $\theta$ and $p$, however its entanglement information only related to $p$. Like the Werner state, when $p> \frac{1}{3}$ it is entangled \cite{G53-Ren-multiple-classify}. 

The second generalized Werner state, which we call the G\uppercase\expandafter{\romannumeral2}-Werner state, is
\begin{equation}\label{eq:10}
	\rho(\theta,\phi)=p\ket{\psi_{\theta,\phi}} \bra{\psi_{\theta,\phi}}+\frac{(1-p)I}{4},
\end{equation}
where, $\ket{\psi_{\theta,\phi}}=\cos \frac{\theta}{2}\ket{00}+e^{i \phi} \sin \frac{\theta}{2}\ket{11}$, $p\in (0,1)$, $\theta \in (0,\pi)$, $\phi \in (0,2\pi)$. The G\uppercase\expandafter{\romannumeral2}-Werner state has three free parameters $\theta$, $p$ and $\phi$, but its entanglement information is only related to $\theta$ and $p$. It is entangled when $p> \frac{1}{(1+2\sin\theta)}$ \cite{G24-Ma}.

Normally, it needs 15 observable operators to reconstruct a 2-qubits state density matrix. However, the number of free parameters of above three quantum states is less than 15, and that of parameters related to entanglement may be even less. In principle, if we can effectively extract and process the entanglement information, we can classify the entanglement of quantum states with the least resource consumption. 

\begin{figure*}
	\centering
	\includegraphics[width=5in]{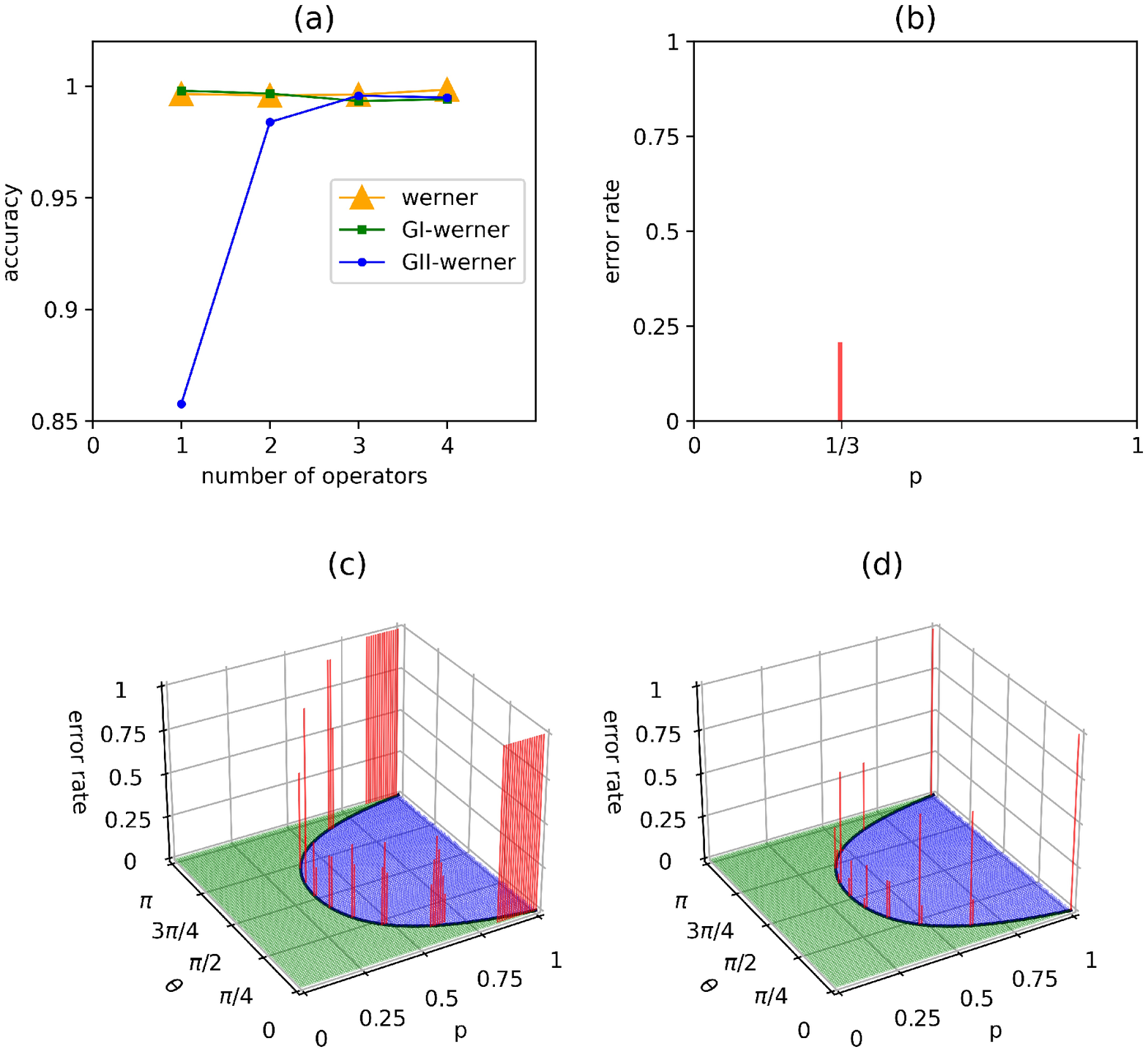}
	\caption{\label{G0G1G2werner acuracys and errors}
		\textbf{(a),}The performance of BCNN for entanglement detection of Werner state, G\uppercase\expandafter{\romannumeral1}-Werner state and G\uppercase\expandafter{\romannumeral2}-Werner state. The accuracy increases with the number of observable operators. \textbf{(b),}The error distribution of the BCNN with 1 observable operators, when entanglement classification is performed on G\uppercase\expandafter{\romannumeral1}-Werner state. When $p> \frac{1}{3}$, the state is entangled. And the errors concentrate on the boundery of entanglement and separability.  \textbf{(c)(d),}The error distribution of the BCNN with 2 and 3 observable operators, when entanglement classification is performed on G\uppercase\expandafter{\romannumeral2}-Werner state. We only drew the  distribution when $\theta=0, 0.1\pi, \cdots, \pi$ for more clear view. The boundery of entanglement and separability is $p=\frac{1}{(1+2\sin\theta)}$. And the errors concentrate on the boundery and the area $\theta=0$ and $\pi$. }
\end{figure*}

We use the BCNN consisting of convolutional paths $(m\in \{1,2,3,4\}; n_1=1, n_2=1)$ and three fully connected layers to classify the entanglement of Werner state, G\uppercase\expandafter{\romannumeral1}-Werner state and G\uppercase\expandafter{\romannumeral2}-Werner state. The convolutional path uesd here has two convolutional layer, and each layer has just one kernel. It can train a observable operator and calculate its average. In practice, based on few observable operators, the BCNN can predict the entanglement of the these quantum states with high accuracy, whitch shown in Fig.\ref{G0G1G2werner acuracys and errors}(a). When classifying the entanglement of the Werner state, the accuracy of the BCNN achieve 99.7\% with only one observable operator $(m=1)$. For the G\uppercase\expandafter{\romannumeral1}-Werner state, FC has achieved 97\% accuracy with two selected observable operators \cite{G53-Ren-multiple-classify} and BCNN can achieve 99.8\% with only one observable operator $(m=1)$. For the G\uppercase\expandafter{\romannumeral2}-Werner state, BCNN can achieve 98.4\% with two observable operators $(m=2)$ and 99.6\% with three observable operators $(m=3)$, which is at the same level with the performance of FC with four selected observable operators \cite{G24-Ma}. (Compared with using a FC with four selected observable operators \cite{G24-Ma}, our results are about the same.)  The error distributions of the BCNN are shown in Fig.\ref{G0G1G2werner acuracys and errors}(b-d). As we can see, the errors are concentrated on the boundary of entanglement and separability. Especially for G\uppercase\expandafter{\romannumeral2}-Werner state, the errors also occur when $\theta=0$ and $\pi$, whitch there are only separable states. The trained observable operators which used to extract the entanglement information can be acquired from the kernels. We show them in TABLE \ref{tab:1} and only keep two decimal places.

\begin{table*}
	\centering
	\caption{\label{tab:1}Trained observable operators}
	\begin{tabular}{cccccccccc}
		\hline
		state & operator number & $X^{(1)}$ & $Y^{(1)}$ & $Z^{(1)}$ & $I^{(1)}$ & $X^{(2)}$ & $Y^{(2)}$ & $Z^{(2)}$ & $I^{(2)}$\\
		\hline
		Werer & 1 & -1.26 & -0.97 & -1.40 & 0.61 & 0.49 & -0.18 & 1.63 & 0.62\\
		\hline
		G\uppercase\expandafter{\romannumeral1}-Werner & 1 & -0.37 & 0.16 & 1.08 & 0.19 & -0.18 & 0.39 & 0.95 & -0.51\\
		\hline
		\multirow{6}{*}{G\uppercase\expandafter{\romannumeral2}-Werner} & \multirow{2}{*}{2} & 0.04 & 0.17 & 0.49 & -1.57 & -0.05 & -0.22 & 1.00 & -0.06\\
		\multirow{6}{*}{} & \multirow{2}{*}{} & -0.05 & -0.09 & 1.39 & 0.07 & 0.02 & 0.22 & 0.97 & 0.17\\
		\\
		\multirow{6}{*}{} & \multirow{3}{*}{3} & 2.71 & -0.27 & 0.50 & 0.56 & 2.40 & -0.96 & 0.48 & -0.53 \\
		\multirow{6}{*}{} & \multirow{3}{*}{} & 0.71 & -1.23 & -1.33 & 1.11 & 0.95 & 0.11 & 0.83 & 0.69 \\
		\multirow{6}{*}{} & \multirow{3}{*}{} & -0.20 & -0.64 & -0.30 & 0.10 & 3.38 & 0.29 & -0.25 & -0.08 \\
		\hline
	\end{tabular}
\end{table*}

\begin{figure*}
	\centering
	\includegraphics[width=6.5in]{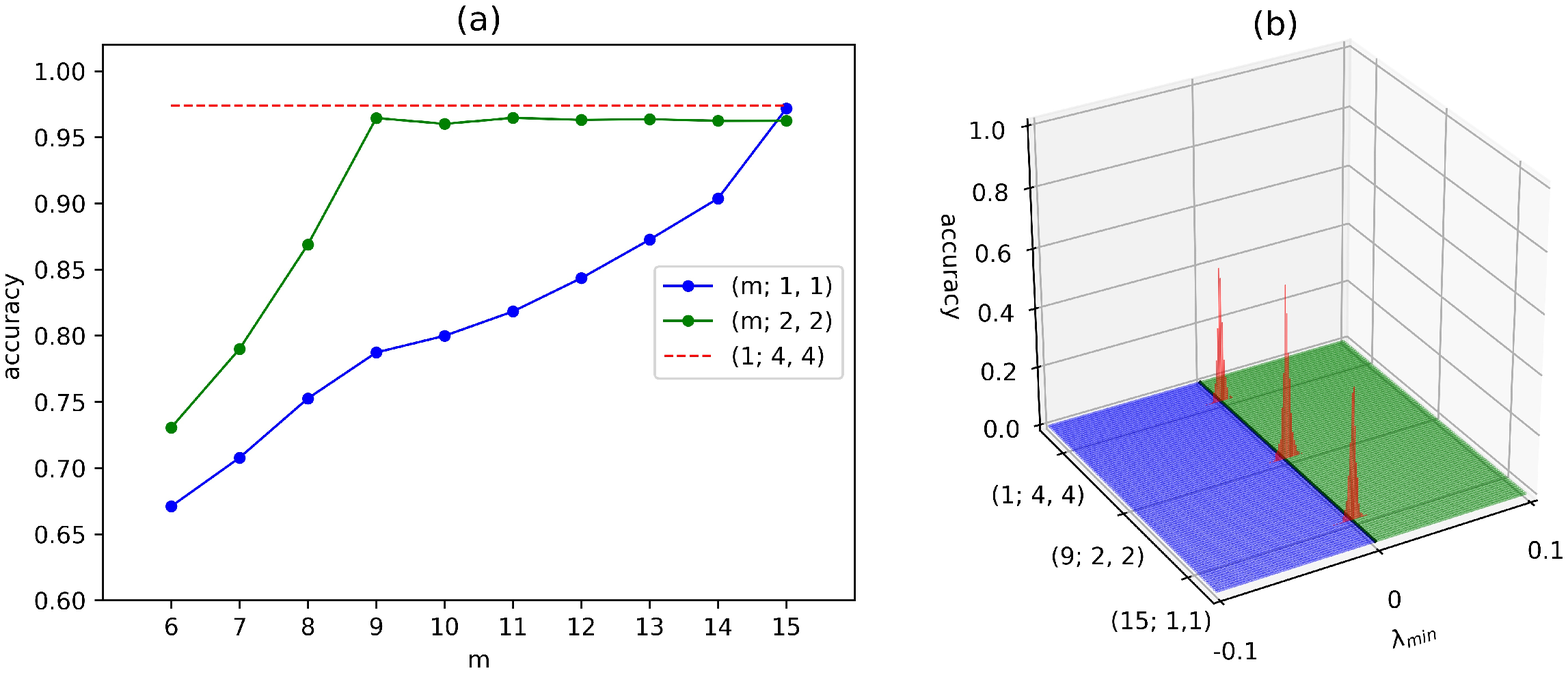}
	\caption{\label{general state accuracy error}
		\textbf{(a),}The performance of BCNN for entanglement detection of 2-qubits general state. The accuracy almost increases linearly with the number of observable operators. \textbf{(b),}The error distribution of the BCNN with 15 observable operators, when entanglement classification is performed on 2-qubits general state. The horizontal axis is the minimum eigenvalue $\lambda_{min}$. The error concentrates on the boundery of entanglement and separability. The error distribution is symmetric about the boundary, so our prediction is unbiased.}
\end{figure*}

Finally, we apply BCNN to classify the entanglement of general 2-qubits state. For the state generation, we adopt the method of $\rho=\frac{\sigma \sigma^+}{tr(\sigma \sigma^+)}$, where $\sigma$ is a random complex matrix and keep the proportion of entangled states and separable states at 1:1. We show the performance of BCNN with three different convolutional path $(m=1; n_1=4, n_2=4)$, $(m\in [6, 15]; n_1=2, n_2=2)$ and $(m\in[6,15]; n_1=1, n_2=1)$ in Fig.\ref{general state accuracy error}. Since the general state is more complicated, we use five fully conneted layers in BCNN. For the structure $(m=1; n_1=4, n_2=4)$, we fix one kernel in each convolutional layer as the identity matrix, and other kernels are still trainable. The outputs of the convolutional path are the averages of 15 observable operators and a constant 1. In this case, the accuracy of BCNN can achieve 97.5\%. For $(m\in [6, 15]; n_1=2, n_2=2)$, we also fix one kernel in each convolutional layer as the identity matrix. Each convolutional path outputs 3 observable operator averages and a constant 1. When $m\geq 9$, the convolutional paths are able to get all the information about the quantum state, and the accuracy of BCNN can be higher than 96.0\%. For the structure $(m\in[6,15]; n_1=1, n_2=1)$, each convolutional path computes the average of just one observable operator. Therefore, with the increase of $m$, the accuracy of this structure raises lowest. And the BCNN still need 15 convolutional paths to extract all quantum state information and its accuracy can achieve 97.2\%. In Fig.\ref{general state accuracy error}(b), we show the error distribution of BCNN with three above convolutional paths when they are just able to extract all the information about quantum state. There only a few errors occur symmetrically around the boundary of entanglement and separability. As long as the structure of the convolutional paths allows all information to be extracted, the BCNN can train appropriate observable operators and detect the entanglement of a general quantum state with high accuracy, which is  comparable to the results in article \cite{G24-Ma}.

\section*{Discussion}
In this work, we prove the observable operator of discrete-level systems is a convolution kernel, which means that the convolutional layer of artificial neural network can accurately calculate the average value of observable operator in quantum state, and that the Hermiticity of the convolutional layer can be maintained with the optimization algorithm based on gradient descent. With the foundation of above, we propose a BCNN, which can obtain well-trained observable operators to efficiently extract entanglement information and classify entanglement. We classify the entanglement of 2-qubits states, and studied the accuracy and the error distribution of BCNN. 

We believe that CNN will be a promising tool for quantum physics. In our work, for Werner state, G\uppercase\expandafter{\romannumeral1}-Werner state and G\uppercase\expandafter{\romannumeral2}-Werner state, it can achieve 99.7\%, 99.8\%, and 98.4\% respectively when the numbers of observable operators are same with that of parameters related to entanglement. It is superior to previous work in reducing resource consumption. For general 2-qubits state, our model still needs 15 observable operators to achieve the accuracy of 97.2\%. In addition, we can extract the trained observable operators from kernels. 
%\textcolor{red}{}
These observable operators can be rewritten as the sum of the orthogonal normalization operators.
We only keep two decimal places of the coefficient and input them into nerual network to test again, and find that the accuracy of the BCNN almost keep the original level.

Since the convolutional layer can accurately calculate the average value of observable operator in quantum state, and this property applies to any dimension discrete-level system. It may be a powerful tool for solving measurement direction of the maximum violation of Bell inequality. We believe this property is likely to be used in other research and can bring new inspiration to the understanding of the relationship between quantum physics and artificial neural networks.

\section*{\label{Methods}METHODS}

\begin{table*}
	\centering
	\caption{\label{tab:2}Adam parameters}
	\begin{tabular}{l|cccccc}
		\hline
		state & operates number & lr & $\beta_1$ & $\beta_2$ & batch size & epoches\\
		\hline
		Werer & 1-15 & 0.001 & 0.9 & 0.99 & 10 & 10\\
		\hline
		G\uppercase\expandafter{\romannumeral1}-Werner & 1-15 & 0.001 & 0.35 & 0.99 & 10 & 10\\
		\hline
		\multirow{3}{*}{G\uppercase\expandafter{\romannumeral2}-Werner}& 1 & 0.001 & 0.5 & 0.9 & 10 & 10\\
		\multirow{3}{*}{}& 2 & 0.001 & 0.9 & 0.99 & 200 & 30\\
		\multirow{3}{*}{}& 3-15 & 0.001 & 0.375 & 0.99 & 10 & 10\\
		\hline
		\multirow{8}{*}{General} & 8 & 0.0003 & 0.325 & 0.825 & 400 & 20\\
		\multirow{8}{*}{} & 9 & 0.0003 & 0.325 & 0.85 & 400 & 20\\
		\multirow{8}{*}{} & 10 & 0.0003 & 0.325 & 0.87 & 400 & 20\\
		\multirow{8}{*}{} & 11 & 0.0003 & 0.325 & 0.9 & 400 & 20\\
		\multirow{8}{*}{} & 12 & 0.0003 & 0.325 & 0.95 & 400 & 20\\
		\multirow{8}{*}{} & 13 & 0.0003 & 0.325 & 0.925 & 400 & 20\\
		\multirow{8}{*}{} & 14 & 0.0003 & 0.325 & 0.925 & 400 & 20\\
		\multirow{8}{*}{} & 15 & 0.0003 & 0.325 & 0.975 & 400 & 20\\
		\hline
	\end{tabular}
\end{table*}

%As a demonstration, we designed a BCNN, depicted in Fig.\ref{branching convolutional neural network}, to detect the entanglement of 2-qubits quantum states. It

The BCNN has several independent convolutional paths and  fully connected layers. A convolutional path has several convolutional layers. The kernel of each convolutional layer represents the transpose of a observable operator acting on the system or subsystem.  Therefore their convolutional layers should not have activation functions and biases. Each convolutional layer outputs all the averages of all combinations of the kernels of its convolutional layers. Then, we take the outputs of the convolutional paths as the input of fully connected layers for entanglement detection. And the structure of convolutional paths and fully connected layers should be adjusted according to the task. In this work, we use the BCNN consists of the convolutional paths $(m; n_1=1,n_2=1)$ and three fully connected layers with the structure $(\alpha,1024,1)$ to detect the entanglement of Wenrer state, G\uppercase\expandafter{\romannumeral1}-Werner state and G\uppercase\expandafter{\romannumeral2}-Werner state. For general 2-qubits state, we test the BCNN consists of one of three different convolutional paths $(m; n_1=4, n_2=4)$, $(m; n_1=2, n_2=2)$ or $(m; n_1=1, n_2=1)$, and five fully connected layers ($\alpha$,1024,1024,1024,1).  The $\alpha$ is the number of the input nodes of the first fully connected layer, which is related to the structure of convolutional paths. The first layer has no activation functions and bias. The final layer's activation function is sigmoid and the loss function is cross entropy \cite{cross-entropy}. For other layers, we take Relu \cite{relu} as the activation function. We use Adam \cite{2014Adam} as our optimizer to make it more likely to cross the saddle point and local minimum. We did not use the default recommended parameters of Adam, 
%\textcolor{red}{}
but adjust them according to the quantum state and the number of convolutional paths. Our adam parameter settings are listed in TABLE \ref{tab:2} for reference.

In training process, we takes state density matrix as the input, and the kernels are initialized to a random Hermitian matrix. According to the features of the quantum state and the structure of convolutional paths, BCNN can automatically find appropriate observable operators for training task. In test process, we can directly calculate the average value of these trained observable operators, and input them into the following fully connected layers to detect entanglement. Of course, based on Eq.(\ref{eq:1}), single convolutional layer can be used to find %\textcolor{red}{}
global observable operators for entanglement detection, but the same task can already be completed by FC \cite{2020Finding}. Therefore, in this article, we will focus on using two convolutional layers to express the product observable operator.

%\section*{DATA AVAILABILITY}
%The data that support the findings of this study are available from the corresponding author upon reasonable request.

%\section*{CODE AVAILABILITY}
%The code that support the findings of this study are available from the corresponding author upon reasonable request.

%\begin{acknowledgments}

%\end{acknowledgments}

%\section*{AUTHOR CONTRIBUTIONS}

%\section*{COMPETING INTERESTS}
%The authors declare no competing interests.

%\nocite{*}

\bibliography{main}% Produces the bibliography via BibTeX.
\bibliographystyle{naturemag}

\end{document}